\documentclass[twocolumn,showpacs,preprintnumbers,amsmath,amssymb]{revtex4}
\usepackage{graphicx}% Include figure files
\usepackage{dcolumn}% Align table columns on decimal point
\usepackage{bm}% bold math

\usepackage{color}
\begin{document}
%\begin{CJK*}{GBK}{}

\preprint{1}

\title{Production and ratio of $\pi$, $K$, $p$ and $\Lambda$ in Pb + Pb collisions at $\sqrt{s_{NN}}$ = 2.76 TeV }% Force line breaks with \\

\author{S. Zhang}
%\email{zhangsong@sinap.ac.cn}
\author{L. X. Han}
\author{Y. G. Ma\footnote{Author to whom all correspondence should be addressed: ygma@sinap.ac.cn}}
\author{J. H. Chen}
\author{C. Zhong}

\affiliation{Shanghai Institute of Applied Physics, Chinese Academy of Sciences, Shanghai 201800, China}

\date{\today}% It is always \today, today,

\begin{abstract}
The particle production and their ratios for $\pi$, $K$, $p$, and
$\Lambda$ are studied in Pb + Pb collisions at $\sqrt{s_{NN}}$ =
2.76 TeV based on a blast-wave model with thermal equilibrium
mechanism. The transverse momentum spectra of the above mentioned
particles at the kinetic freeze-out stage are discussed. The
modification of the inverse slope of pion transverse momentum
spectrum due to resonance decay has also been investigated. In
addition, we found that the anti-particles to particles ratio as
well as kaons to pions ratio agree with the data by the LHC-ALICE
Collaboration reasonably well, while the $p/\pi$ ratio is
overestimated by a factor of 1.5, similar to those from other
thermal model calculations. It is found that the ratios of $p/\pi$
and $K/\pi$ are dominated by the radial flow but slightly affected
by the baryon chemical potential. Our study thus constrains the
parameters at the chemical and kinetic freeze-out stages within the
framework of thermal model in Pb + Pb collisions at $\sqrt{s_{NN}}$
= 2.76 TeV, and will help better understand the properties of the
dense and hot matter created in high-energy heavy-ion collisions at
freeze-out stage.
\end{abstract}

\maketitle

%%%%%%%%%%%%%%%%%%%%%%%%%%%
\section{Introduction}

Ultra-relativistic heavy-ion collisions opens a window for studying
the properties of the Quark-Gluon Plasma (QGP) which was predicted
by quantum chromodynamics (QCD)~\cite{QCD-QGP}. This exotic matter
is believed to be produced in the early stage of central Au + Au
collisions at the top energy in the Relativistic Heavy-Ion Collider
(RHIC) at Brookhaven National Laboratory~\cite{RHICWithePaper}. The
sufficient experimental evidences~\cite{RHIC-SQGP} support that the
new hot and dense QCD matter is not an ideal gas but instead a
strongly interacting dense partonic matter named as sQGP under
extreme temperature and energy density. The collective properties of
the exotic matter created at RHIC can be investigated through
transverse momentum ($p_{T}$) distribution and elliptic flow of
identified particles, and so far people have found that this matter
behaves as a nearly ideal fluid~\cite{RHIC-pLiquid}. Recently,
experimental results in Pb + Pb collisions at $\sqrt{s_{NN}}$ = 2.76
TeV in the Large Hadron Collider (LHC) are also reported by  the
ALICE
Collaboration~\cite{ALICE-chDen,ALICE-chCDepen,ALICE-chRAA,ALICE-chPCDepen}.
This provides another opportunity to investigate the bulk properties
of the exotic QCD matter as an expanding fireball created in
heavy-ion collisions at a higher energy, such as its baryon chemical
potential $\mu_{B}$, chemical freeze-out temperature $T_{ch}$, and
kinetic freeze-out temperature $T_{kin}$ as well as radial expansion
velocity $\langle \beta_{T} \rangle$, etc.

Relativistic hydrodynamics and thermal models are very successful in
describing particle productions at the freeze-out stage. The viscous
hydrodynamic model VISH2+1~\cite{VISH2} has successfully described
the transverse momentum distributions of $\pi$ and $K$. Similar
model named HKM~\cite{HKM} coupling with UrQMD for the hadronic
scattering stage can also reproduce the yields and distributions of
particles such as $\pi$, $K$, and $p$. The EPOS model~\cite{EPOS}
aims at describing complete transverse momentum distributions of
particles within the same dynamical picture. A multiphase transport
(AMPT) model has been reconfigurated to reproduce the $p_{T}$
distribution of charged particles as well as their elliptic flow in
Pb + Pb collisions at $\sqrt{s_{NN}}$ = 2.76 TeV~\cite{AMPT-LHC}.
Besides the studies from hydrodynamics or transport models mentioned
about, the thermal model has also successfully described the
production of particles in heavy-ion collisions with a few
parameters such as the chemical freeze-out temperature, the baryon
chemical potential, and the fireball volume~\cite{Thermal-PMB}. From
particle ratios, the thermal model~\cite{Thermal-EQ} can be used to
obtain the chemical freeze-out properties, such as the chemical
freeze-out temperature $T_{ch}$ as well as the baryon ($\mu_B$) and
the strangeness ($\mu_S$) chemical potential. By fitting the
transverse momentum distribution, the blast-wave model~\cite{BLWave}
has often been used to extract the kinetic freeze-out properties
such as the kinetic freeze-out temperature $T_{kin}$ and the radial
flow velocity $\langle \beta_{T} \rangle$. These thermal models have
also been applied in experimental
analysis~\cite{ALICE-chPCDepen,STARSYS-Spectra} to study the
chemical and kinetic freeze-out properties. Retie\`ere and
Lisa~\cite{BLWave-Fabrice} have explored in detail an analytic
parametrization of the freeze-out configuration and investigated the
spectra, the collective flow, and the HBT correlation of hadrons
produced in head-on collisions at top RHIC energy. In addition, the
DRAGON model~\cite{DRAGON} and the THERMINATOR2~\cite{THERMINATOR}
model have also been developed to study the phase-space distribution
of produced hadrons at freeze-out stage.

Due to the complicated initial condition and dynamical evolution in
heavy-ion collisions, it is very likely that the particle production
can be explained with different parameter sets of temperature,
chemical potential, and radial flow.
The values of these parameters give the acceptable range of the
system properties at freeze-out stage. The results in this paper
come from the fitting based on the blast-wave model with thermal
equilibrium mechanism. Transverse momentum ($p_{T}$) distributions
of charged hadrons ($\pi$, $K$, and $p$) and $\Lambda$ hyperon in Pb
+ Pb collisions at $\sqrt{s_{NN}}$ = 2.76 TeV are presented. The
effect from resonance decay to the $p_T$ spectrum of pions will also
be discussed. The inverse slope parameter of the $p_{T}$ spectrum is
extracted for each particle species at various centralities. The
multiplicities of anti-particles and particles become similar by
tuning baryon chemical potential to about 0.1 MeV at LHC energy. The
inclusive yields of charged hadrons and hyperons normalized to the
pion yield in $0-5$\% centrality are compared with the experimental
results. From the transverse momentum dependence of mixed ratios of
$p/\pi$ and $K/\pi$, the radial flow effect on the mass ordering of
$p_{T}$ distribution is investigated. The calculated results agree
pretty well with the experimental measurements by the LHC-ALICE
Collaboration, and a reasonable range of parameters at the chemical
and kinetic freeze-out stages is discussed for the thermalized
system at LHC energy.

%%%%%%%%%%%%%%%%%%%%%%%%%%%%%%%%
\section{blast-wave model with thermal equilibrium mechanism}

As discussed above, thermal model can describe particle yield by
adjusting parameters such as the chemical freeze-out temperature
$T_{ch}$, the baryon chemical potential $\mu_{B}$, the strangeness
chemical potential $\mu_{S}$, and the system volume $V$. On the
other hand, one can extract these quantities at chemical freeze-out
stage through particle ratios. The particle density of species
\textit{i} can be expressed as~\cite{Thermal-EQ, Thermal-PMB,
DRAGON}
\begin{widetext}
\begin{eqnarray}
n_{i}(T_{ch},\mu_{B},\mu_{S})
&=& g_{i}\int \frac{d^{3}p}{(2\pi)^{3}}\left[\text{exp}\left(\frac{\sqrt{p^{2}+m_{i}^{2}}-\left(\mu_{B}B_{i}+\mu_{S}S_{i}\right)}{T_{ch}}\right)\mp1\right]^{-1}\nonumber\\
&=&I\left(g_i,\frac{m_i}{T_{ch}}\right)\sum_{n=1}(\pm 1)^{n+1}\exp{\left(n\frac{\left(\mu_{B}B_{i}+\mu_{S}S_{i}\right)}{T_{ch}}\right)},\nonumber\\
I\left(g_i,\frac{m_i}{T_{ch}}\right)&=&g_{i}\int\frac{d^{3}p}{(2\pi)^{3}}\left[\sum_{n=1}(\pm
1)^{n+1}\exp{\left(-n\frac{\sqrt{p^{2}+m_{i}^{2}}}{T_{ch}}\right)}\right],
\label{eq:ni}
\end{eqnarray}
\end{widetext}
with the upper (lower) sign for bosons (fermions) and $g_{i}$ being
the degeneracy factor. Assuming that the chemical equilibrium
condition is satisfied, Eq.~(\ref{eq:ni}) essentially determines the
fraction of particle species \textit{i}. Within the framework of the
blast-wave model, the fireball created in high-energy heavy-ion
collisions is assumed to be in local thermal equilibrium and expands
at a four-component velocity $u_{\mu}$. The phase-space distribution
of hadrons emitted from the expanding fireball can be expressed as a
Wigner function~\cite{BLWave-Fabrice,DRAGON,THERMINATOR}
\begin{widetext}
\begin{eqnarray}
S(x,p)d^{4}x = \frac{2s+1}{(2\pi)^{3}}m_{t}\text{cosh}(y-\eta)\text{exp}\left(-\frac{p^{\mu}u_{\mu}}{T_{kin}}\right)\Theta(1-\tilde{r}(r,\phi))H(\eta)\delta(\tau-\tau_{0})d\tau\tau d\eta rdrd\phi,
\label{eq:BLWWigner}
\end{eqnarray}
\end{widetext}
where $s$, $y$, and $m_{t}$ are respectively the spin, rapidity, and
transverse mass of the hadron, and $p_{\mu}$ is the four-component
momentum. Equation~(\ref{eq:BLWWigner}) is formulated in a Lorentz
covariant way, $r$ and $\phi$ are the polar coordinates, and $\eta$
and $\tau$ are the pseudorapidity and the proper time, respectively.
$\tilde{r}$ is defined as
\begin{equation}
\tilde{r} =
\sqrt{\frac{(x^{1})^{2}}{R^{2}}+\frac{(x^{2})^{2}}{R^{2}}},
\label{eq:tilde_r}
\end{equation}
with ($x^{1}, x^{2}$) standing for the coordinates in the transverse
plane and $R$ being the average transverse radius. The kinetic
freeze-out temperature $T_{kin}$ and the radial flow parameter
$\rho_{0}$ are important in determining the transverse momentum
spectrum, with the latter affecting the four-component velocity
field. Since we are only interested in the $p_T$ spectrum at
mid-rapidity, the pseudorapidity distribution $H(\eta)$ is not
important. The $p_{T}$ spectrum can then be written as
\begin{equation}
\frac{dN}{2\pi p_{T}dp_{T}} = \int S(x,p)d^{4}x,
\label{eq:pTSpectra}
\end{equation}
and the fraction of particle species $i$ and its phase-space
distribution can be calculated from Eqs.~(\ref{eq:ni}) and
(\ref{eq:pTSpectra}).

%%%%%%%%%%%%%%%%%%%%%%%%%%%%%%
\section{Transverse momentum spectra}

\begin{figure*}
\includegraphics[width=15cm]{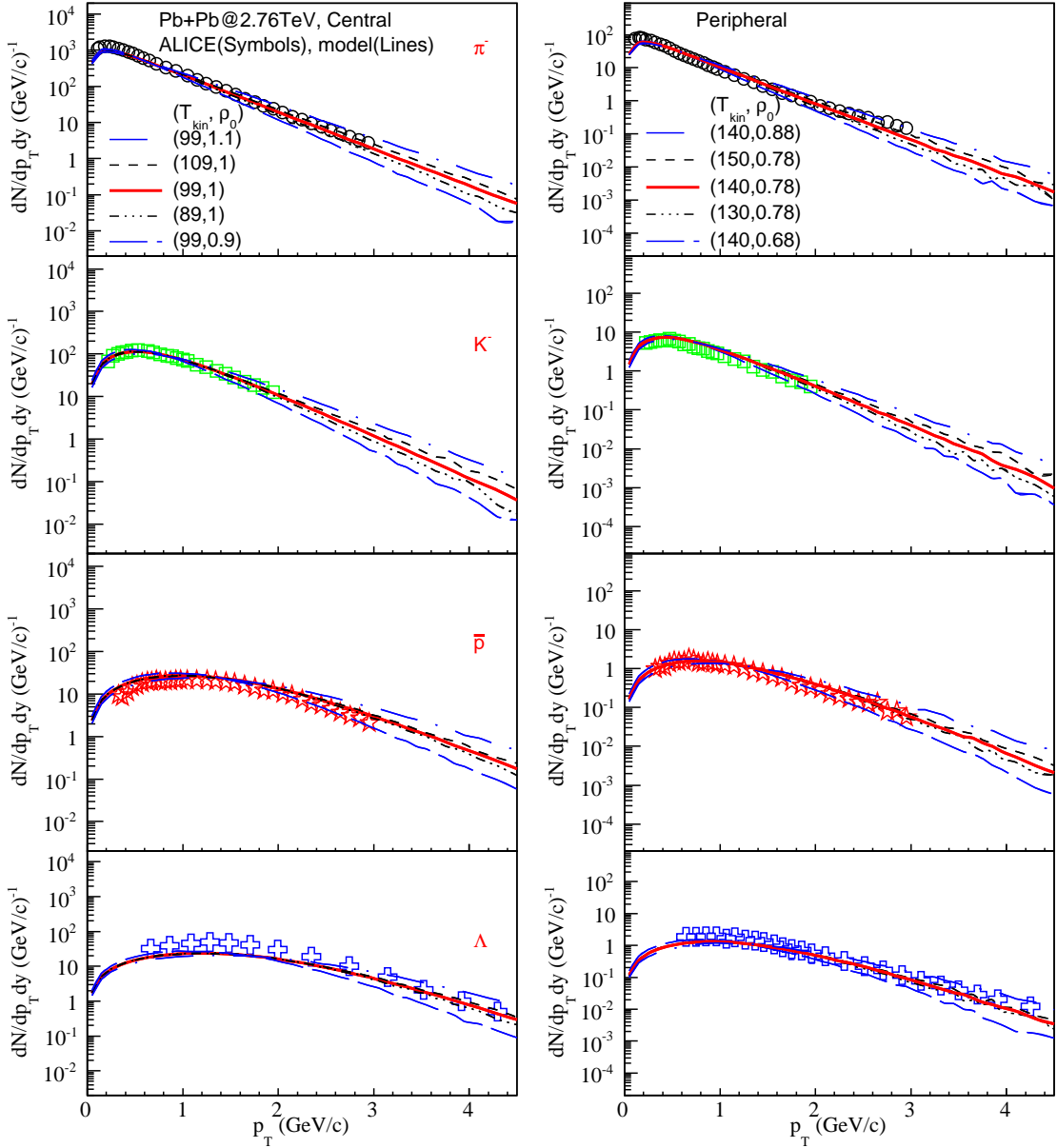}
\caption{\label{fig:spectra} (Color online) Transverse momentum
($p_{T}$) distribution of $\pi$, $K$, $p$, and $\Lambda$ in central
and peripheral collisions. Lines: model calculations with
$\mu_{B}=0.1$ MeV, $\mu_{S}=0.001$ MeV, and $T_{ch}=160$ MeV for
central collisions and $T_{ch}=150$ MeV for peripheral collisions;
Symbols: ALICE
data~\cite{ALICE-chPCDepen,ALICE-Muller,ALICE-IP-pp-PbPb,ALICE-STRANGE-HotQ},
collision centrality $0-5\%$ (left column) and $70-80\%$ (right
column) for $\pi^{-}$, $K^{-}$, and $p$, and $0-10\%$ (left column)
and $60-80\%$ (right column) for $\Lambda$. $T_{kin}$ is in MeV and
$\rho_0$ is dimensionless.}
\end{figure*}

The collective properties of the hot and dense matter created in
ultra-relativistic heavy-ion collisions at freeze-out stage can be
studied through transverse momentum ($p_{T}$) distributions of
identified particles. Figure~\ref{fig:spectra} shows the $p_{T}$
distributions of $\pi$, $K$, $p$, and $\Lambda$ in central and
peripheral Pb + Pb collisions at $\sqrt{s_{NN}}$ = 2.76 TeV by using
the blast-wave model with thermal equilibrium mechanism. The
parameters of kinetic temperature ($T_{kin}$) and radial flow
parameter ($\rho_{0}$) used in the calculation are also shown in
Fig.~\ref{fig:spectra}. The spectra become stiffer with increasing
$\rho_{0}$ and $T_{kin}$, and the results imply that the transverse
momentum distribution is more sensitive to the radial flow parameter
than to the kinetic freeze-out temperature. The ranges of the
chemical and kinetic temperature ($T_{ch}$ and $T_{kin}$) as well as
the chemical potential ($\mu_{B}$ and $\mu_{S}$) are consistent with
those from other model calculations~\cite{VISH2,Thermal-PMB} and the
experimentally estimated values~\cite{ALICE-chPCDepen,
STAR-HBT-BLWave2005}. The radial flow
\begin{equation}
\langle \beta_{T} \rangle = \int
\text{arctanh}\left(\rho_{0}\frac{r}{R}\right)rdr/\int rdr
\end{equation}
is related to the maximum flow rapidity
\begin{equation}
\rho = \tilde{r}\left[\rho_{0}+\rho_{a}\text{cos}(2\phi)\right].
\end{equation}
The results are independent of $\rho_{a}$ as after integration the
$\phi$ dependence is averaged out. The parameter $\rho_{0}$ from our
analysis is comparable to that extracted from the experimental
data~\cite{STAR-HBT-BLWave2005,ALICE-chRAA,ALICE-chPCDepen,ALICE-Muller}
except for $\Lambda$ in central collisions ($0-10\%$). In addition,
we can see that the effect from the radial flow parameter $\rho_0$
on the spectra is more significant than the kinetic freeze-out
temperature $T_{kin}$. The results show that the inverse slopes
$T_{loc}$ increase with $\rho_0$  in the range of $\rho_0$ from 0.9
to 1.1, which is consistent with the previous calculation using the
blast-wave model~\cite{BLWave-Fabrice}.

\begin{figure*}
\includegraphics[width=15.2cm]{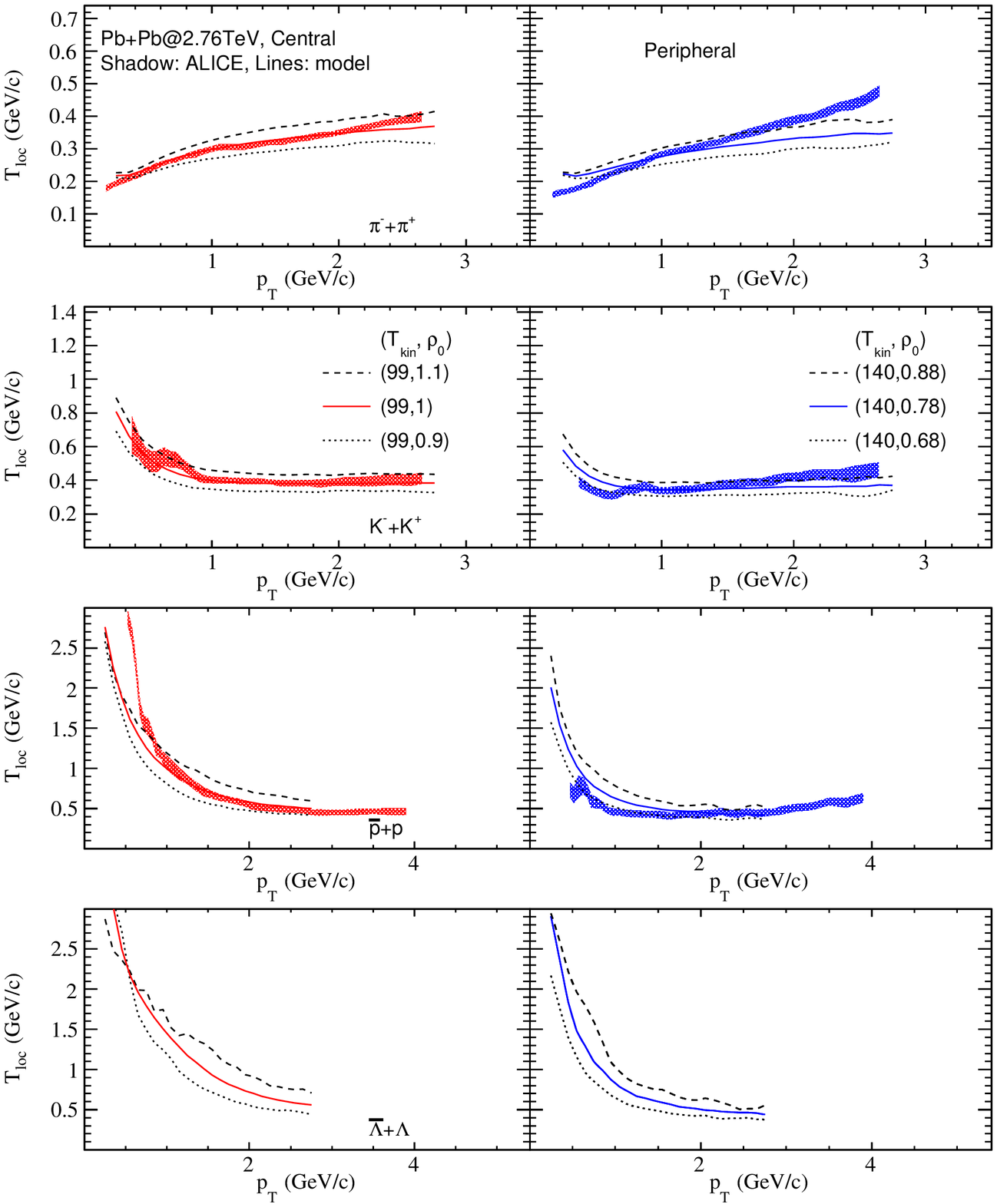}
\caption{\label{fig:Tloc} (Color online) Local slopes of the $p_{T}$
distributions for $\pi$, $K$, $p$, and $\Lambda$ including their
anti-particles as a function of $p_{T}$ in central and peripheral
collisions. Lines: model calculations with $\mu_{B}=0.1$ MeV,
$\mu_{S}=0.001$ MeV, and $T_{ch}=160$ MeV for central collisions and
$T_{ch}=150$ MeV for peripheral collisions; Shadow: the ALICE
data~\cite{ALICE-chPCDepen} for collision centrality $0-5\%$ (left
column) and $70-80\%$ (right column). $T_{kin}$ is in MeV and
$\rho_0$ is dimensionless.}
\end{figure*}

The properties of a thermalized system can be extracted from the
spectra of identified particles. A rough description of spectra
slope can be obtained by an exponential or a power-law fit. From
Fig.~\ref{fig:spectra}, we can see that the spectra of identified
particles become stiffer from peripheral collisions to central
collisions. Experimental data have demonstrated that there is no
obvious onset of the power-law tail at high $p_{T}$ as observed in
pp collisions~\cite{LHC-pp}, but in central collisions the spectra
keep an almost exponential shape in a wide $p_{T}$ range. The
transverse momentum dependence of the slope parameters can display
more clearly the trend of the spectra changing from the exponential
pattern to the power-law one. Figure~\ref{fig:Tloc} shows the local
inverse slope $T_{loc}$ of the spectra as a function of $p_{T}$,
which is calculated by fitting the spectra with the following
function~\cite{ALICE-chPCDepen}
\begin{equation}
\frac{1}{p_{T}}\frac{dN}{d_{p_{T}}}\propto e^{-p_{T}/ T_{loc}}.
\label{eq:pTSlopeFit}
\end{equation}
It is seen that the inverse slopes of identified particles $p_T$
spectra have a similar trend of $p_{T}$ dependence to the
experimental results~\cite{ALICE-chPCDepen}. The inverse slopes of
the $p_T$ spectra for $K$, $p$, and $\Lambda$ decrease with the
increasing transverse momentum, and this is more obvious for central
collisions. In addition, they don't change with increasing $p_{T}$
above a certain value of the transverse momentum, which is about 1
GeV/c for $K$ and 2 GeV/c for $p$ and $\Lambda$. It is also seen
that $T_{loc}$ for protons and kaons converges to a similar value of
0.45 GeV/c at high $p_T$. In peripheral collisions from the
ALICE~\cite{ALICE-chPCDepen} data, a modest increase of $T_{loc}$ is
seen at the highest $p_{T}$, showing the onset of a power-law
behavior. The above experimental results thus indicate that the
system becomes more thermalized with sufficient particle interaction
in central collisions in comparison with peripheral collisions.

\begin{figure*}
\includegraphics[width=15.2cm]{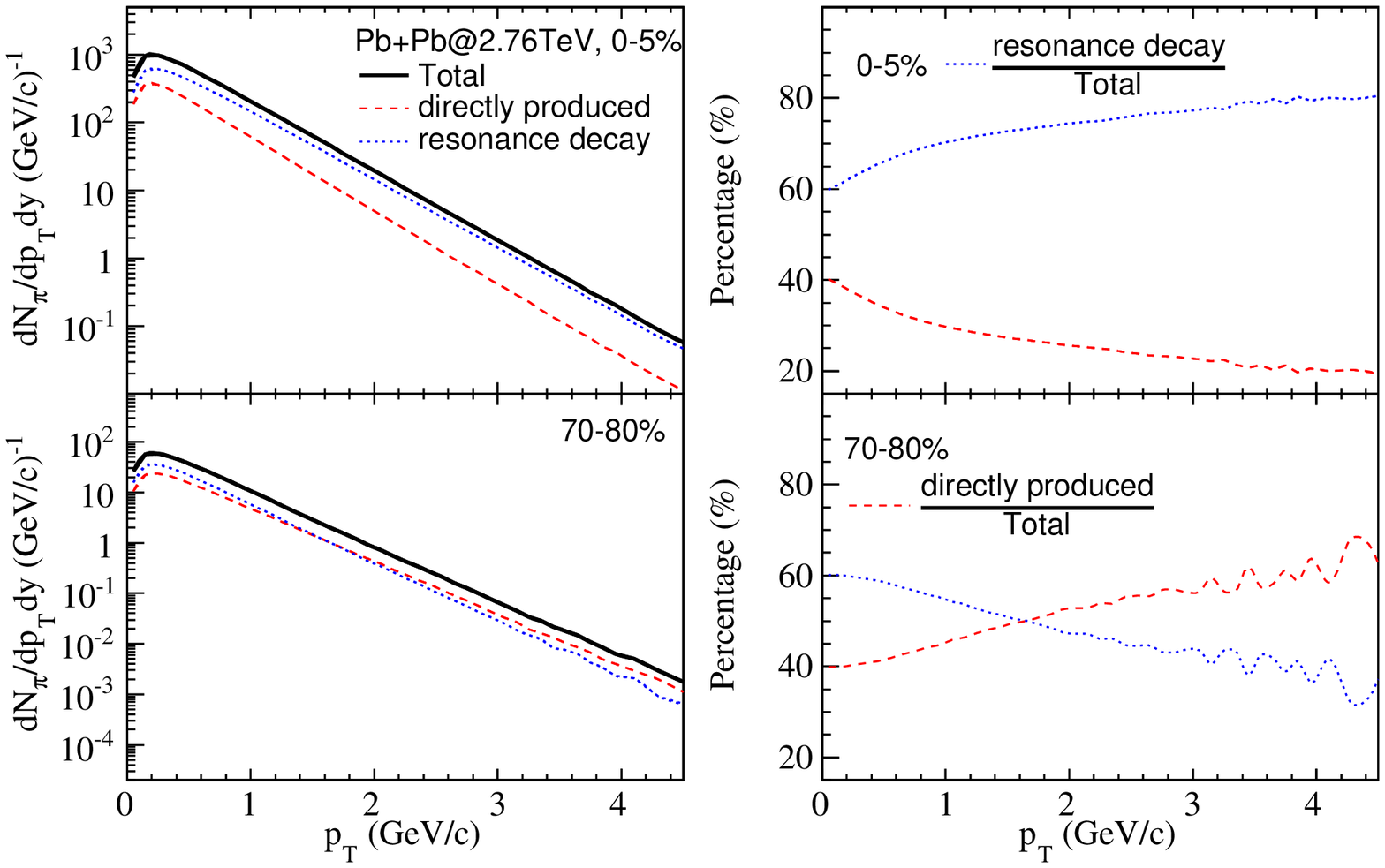}
\caption{\label{fig:spectraDecay} (Color online) Comparing the pion
$p_T$ spectrum from direct production and resonance decay from model
calculations with $(T_{ch}, \mu_B, \mu_S, T_{kin}, \rho_0)=(160,
0.1, 0.001, 99, 1)$ (all in MeV except that $\rho_0$ is
dimensionless) for central collisions and $(T_{ch}, \mu_B, \mu_S,
T_{kin}, \rho_0)=(150, 0.1, 0.001, 140, 0.78)$ for peripheral
collisions. Left: Pions spectra from direct production and total
yield in central and peripheral collisions; Right: The ratios of
pions from resonance decay and direct production to the total yield
as a function of $p_{T}$ in central and peripheral collisions.}
\end{figure*}

For pions, however, the inverse slopes increase with $p_{T}$ in both
central and peripheral collisions, opposite to the trend observed
for protons and kaons. At high $p_{T}$, the power-law rise is more
suppressed in central collisions in comparison with peripheral ones.
The calculated results are consistent with the experimental ones
except at high $p_{T}$ ($p_T>1.5$ GeV/$c$) in peripheral collisions.
The pions from resonances decay contribute to the distribution at
low transverse momenta, and this might be the reason why a different
trend of pion spectrum compared to protons and kaons is
observed~\cite{EPOS,BLWave,STARSYS-Spectra}. The contribution of
resonance decay to the total pion yield is estimated in this
calculation by distinguishing directly produced pions from the
fireball and pions from resonance decay, and they are compared in
Fig.~\ref{fig:spectraDecay}. In central collisions (left top), the
yield of pions from resonance decay is higher than those from direct
production, and this is more pronounced in the top of the right
figures, i.e., the fraction of the contribution from resonance decay
in the total yield grows with increasing $p_{T}$ from $60\%$ to
$80\%$. In peripheral collisions (left and right bottom), however,
this fraction decreases with increasing $p_{T}$ from $60\%$ to
$40\%$, and the directly produced pion is dominant above $p_{T}\sim$
2 GeV/c. This implies that the contribution of the resonance decay
is dominant in central collisions but only important at low $p_{T}$
in peripheral collisions.

%%%%%%%%%%%%%%%%%%%%
\section{Particle ratio analysis}

\begin{table*}
\caption{ \label{tab:antimatter2matter} Ratios of anti-particles to
particles for $\pi$, $K$, $p$, and $\Lambda$ in central and
peripheral collisions. Results from model calculations with
$T_{kin}=99$ MeV and $\rho_{0}=1$ for central collisions and
$T_{kin}=140$ MeV and $\rho_{0}=0.78$ for peripheral collisions are
compared with the ALICE data~\cite{ALICE-chPCDepen} for collision
centrality $0-5\%$ and $70-80\%$. }
\begin{ruledtabular}
\begin{tabular}{lllll}
($T_{ch}$, $\mu_B$, $\mu_S$) (MeV) & $\pi^{-}$/$\pi^{+}$ & $K^{-}$/$K^{+}$ & $\bar{p}$/p & $\bar{\Lambda}$/$\Lambda$\\
\hline
Centrality (0-5\%), ALICE & 0.998 & 1.00 & 0.971 & -\\
%\hline
(160, 0.1, 0.001) & 0.99 & 0.99 & 1.00 & 0.99\\
%\hline
(150, 0.1, 0.001) & 0.99 & 0.99 & 1.00 & 0.98\\
%\hline
(170, 0.1, 0.001) & 0.99 & 0.99 & 1.00 & 1.00\\
%\hline
(160, 0.1, 0.1)     & 0.99 & 0.99 & 1.00 & 0.99\\
%\hline
(160, 10, 0.001)  & 0.99 & 0.99 & 0.88 & 0.88\\
\hline
Centrality (70-80\%), ALICE & 0.994 & 1.00 & 1.033 & -\\
(150, 0.1, 0.001) & 0.997 & 0.995 & 1.003 & 0.997\\
(140, 0.1, 0.001) & 1.000 & 0.998 & 1.006 & 1.018\\
(160, 0.1, 0.001) & 0.995 & 0.993 & 0.994 & 0.994\\
(150, 0.1,   0.1) & 0.997 & 0.992 & 0.992 & 0.997\\
(150, 10,  0.001) & 0.998 & 0.994 & 0.879 & 0.877

\end{tabular}
\end{ruledtabular}
\end{table*}

The estimate of the baryon chemical potential $\mu_{B}$ gives the
value of about zero from the similar multiplicity of anti-particles
and particles measured by the ALICE
Collaboration~\cite{ALICE-chPCDepen}. The anti-particles to
particles ratio can be approximately deduced from Eq.~(\ref{eq:ni})
by neglecting the second- and higher-order terms, i.e.,
\begin{eqnarray}
\frac{\bar{n}_i}{n_i} &\approx& \exp{\left(\frac{\mu_B(\bar{B}_i-B_i)+\mu_S(\bar{S}_i-S_i)}{T_{ch}}\right)}\nonumber\\
&=&\exp{\left(-2\frac{\mu_B|B_i|+\mu_S|S_i|}{T_{ch}}\right)}.
\label{eq:ratio-antim-m}
\end{eqnarray}
From Eq.~(\ref{eq:ratio-antim-m}), it can be seen that the
multiplicity ratio of anti-particle to particle is affected by the
chemical properties of the bulk matter such as $T_{ch}$, $\mu_B$,
and $\mu_S$. Table.~\ref{tab:antimatter2matter} gives the
multiplicity ratios of anti-particles to particles from different
parameter sets used in the calculation. It is seen that there is
minor effect from the chemical freeze-out temperature $T_{ch}$ and
the strangeness chemical potential $\mu_S$ on the ratio. However,
the baryon chemical potential $\mu_B$ dominates the ratio of
anti-baryon to baryon ($\bar{p}/p$ and $\bar{\Lambda}/\Lambda$). It
is obviously seen that the ratios are compatible with unit for
centralities of $0-5\%$ and $70-80\%$ with the chemical potential
$\mu_B$ and $\mu_S$ close to zero, consistent with the experimental
observation~\cite{ALICE-chPCDepen}. This means that the experimental
results can be well described by thermal equilibrium mechanism, and
this implies that the hot and dense matter created at LHC energy has
nearly equal amount of matter and anti-matter with the chemical
potential close to zero.

\begin{figure*}
\includegraphics[width=8.5cm]{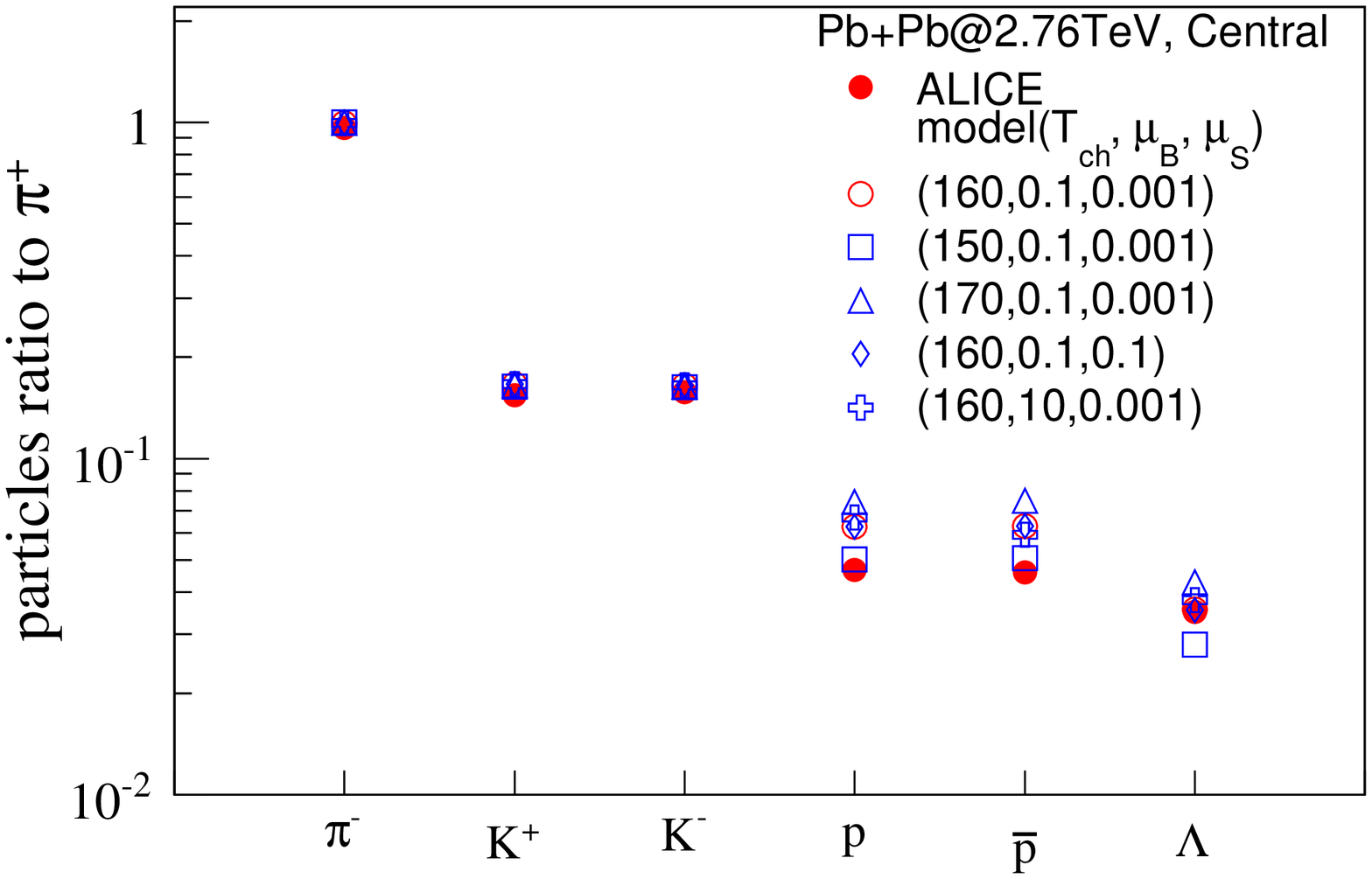}
\caption{\label{fig:ratioToPion}  (Color online) Ratio of inclusive
yields of particles to $\pi^{+}$ compared with the LHC-ALICE data
for central ($0-5\%$) collisions~\cite{ALICE-Muller,
ALICE-raioToPion}. $T_{ch}$, $\mu_B$, and $\mu_S$ are all in MeV.}
\end{figure*}

The accurate ratio of inclusive yields of particles to those of
$\pi^{+}$ can be deduced from Eq.~(\ref{eq:ni}) as done in
Eq.~(\ref{eq:ratio-antim-m})
\begin{eqnarray}
\frac{n_i}{n_{\pi^{+}}}
&=\frac{I\left(g_i,m_i/T_{ch}\right)}{I\left(g_{\pi^{+}},m_{\pi^{+}}/T_{ch}\right)}\exp\left(\frac{\mu_BB_i+\mu_SS_i}{T_{ch}}\right).
\label{eq:ratio-particleTopion-m}
\end{eqnarray}
The ratio of inclusive particle yields to that of $\pi^{+}$ is thus
determined by $T_{ch}$, $\mu_B$, and $\mu_S$ in addition to the
intrinsic parameters of mass $m_i$ and the degeneracy factor $g_i$.
Figure~\ref{fig:ratioToPion} shows particle yield ratios of hadrons
to $\pi^{+}$ in $0-5\%$ centrality and they are compared with the
experimental results~\cite{ALICE-Muller, ALICE-raioToPion}. Within
the range of the chemical freeze-out temperature $T_{ch}$ and the
chemical potentials $\mu_{B}$ and $\mu_{S}$ as shown in
Fig.~\ref{fig:ratioToPion}, particle yield ratios of most hadrons to
$\pi^{+}$ in this calculation are consistent with those in
Ref.~\cite{Thermal-PMB}. However, the thermal model can not
reproduce the yield of protons ($T_{ch}=150$ MeV) and $\Lambda$
hyperons ($T_{ch}=160$ MeV) with the same parameters, and it calls
for more theoretical works for particle production mechanism, which
is beyond what we have in the present paper.

\begin{figure*}
\includegraphics[width=15cm]{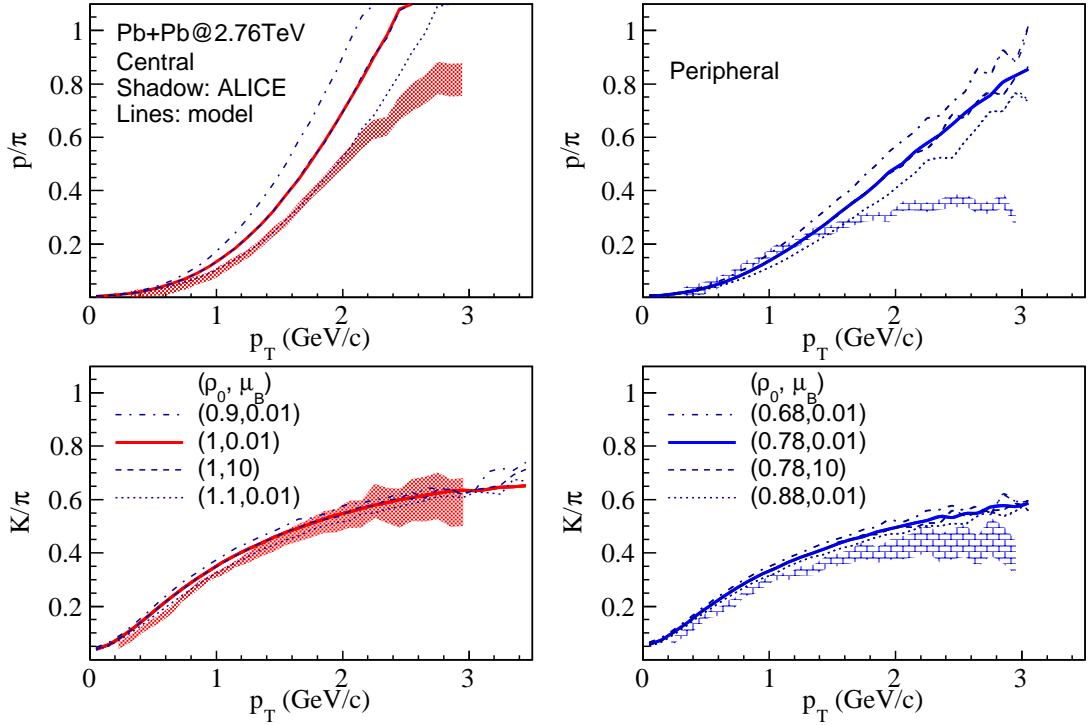}
\caption{\label{fig:ratiopT}  (Color online) Ratios of $p/\pi$
(stands for $(p+\overline{p})/(\pi^{+}+\pi^{-}))$ and $K/\pi$
(stands for $(K^{+}+K^{-})/(\pi^{+}+\pi^{-}))$ as a function of
$p_{T}$ in central and peripheral collisions. Lines: model
calculations with $\mu_S = 0.001$ MeV, $T_{ch} = 160$ MeV and
$T_{kin} = 99$ MeV for central collisions, and $T_{ch} = 150$ MeV
and $T_{kin} = 140$ MeV for peripheral collisions; Shadow: the ALICE
data~\cite{ALICE-chPCDepen} for collision centrality $0-5\%$ (left
column) and $70-80\%$ (right column). $\rho_0$ is dimensionless and
$\mu_B$ is in MeV.}
\end{figure*}

The transverse momentum dependence of mixed ratio of $p/\pi$ (stands
for $(p+\overline{p})/(\pi^{+}+\pi^{-})$) and $K/\pi$ (stands for
$(K^{+}+K^{-})/(\pi^{+}+\pi^{-})$) will be affected by both chemical
freeze-out properties and kinetic freeze-out properties. From the
above discussion, it is known that the slope of particle spectra and
their ratios are mainly determined by the radial flow parameter
$\rho_0$ and the baryon chemical potential $\mu_B$, respectively. It
will be interesting to investigate these ratios in different ranges
of $\rho_0$ and $\mu_B$ used above. The ratios of $p/\pi$ and
$K/\pi$ as a function of $p_{T}$ are shown in
Fig.~\ref{fig:ratiopT}. It is seen that there is no significant
effect from $\mu_B$ on the particle ratio, while a large radial flow
parameter $\rho_0$ will reduce the ratios, especially for $p/\pi$.
The effect from $\rho_0$ as well as the increasing trend of the
ratios with transverse momentum are intrinsic features of
hydrodynamical models, where heavier particles are pushed to higher
$p_{T}$ by the collective motion of radial flow. The increasing
trend of both ratios with $p_{T}$ is more pronounced in central
collisions and this is consistent with the LHC-ALICE
results~\cite{ALICE-chPCDepen}. However, the ratios from our
calculations are higher than the experimental results, except for
$K/\pi$ in central collisions. For the $p/\pi$ ratio, the result
from our calculations and those by other models~\cite{HKM,
Hydro-Flow-RHIC-LHC} overestimate the value measured by the
LHC-ALICE Collaboration~\cite{ALICE-chPCDepen}.

\section{Summary}

The particle yield and their ratios of $\pi$, $K$, $p$, and
$\Lambda$ in Pb + Pb collisions at $\sqrt{s_{NN}}$ = 2.76 TeV have
been investigated based on the blast-wave model with thermal
equilibrium mechanism. A reasonable range of the parameters at
chemical and kinetic freeze-out stages in our calculation is
selected to study the thermalized system at LHC energy. The
transverse momentum spectra demonstrate some thermal and dynamical
properties of the system, such as the radial flow, the chemical and
kinetic temperatures, and the baryon (strangeness) chemical
potential. Similar to the early findings, the slope of spectra is
dominated by the radial flow parameter $\rho_0$, while the
anti-particles to particles ratio is essentially controlled by the
baryon chemical potential, and the transverse momentum dependence of
mixed ratios of $p/\pi$ and $K/\pi$ are only affected by $\rho_0$.
It is found that the slopes of $K$ and $p$ spectra becomes
independent of $p_T$, which is about 0.4 GeV/c for $K$ and 0.5 GeV/c
for $p$, when the transverse momentum is above a certain value,
which is about 1 GeV/c for $K$ and 2 GeV/c for $p$, and this
indicates an exponential shape of the $p_T$ spectra at high $p_T$.
The modification of the inverse slope of transverse momentum spectra
for pions due to resonance decay has also been investigated. The
anti-particles to particles ratios are compatible with unity in
central and peripheral collisions, which is consistent with the LHC
results. This implies that the baryon chemical potential is almost
zero at LHC energy. The inclusive yields of particles normalized to
$\pi^{+}$ are comparable to those measured by the LHC-ALICE
Collaboration but the $p/\pi$ ratio is overestimated by a factor of
1.5 even though it is similar to those from other thermal model
calculations. The ratios of $p/\pi$ and $K/\pi$ as a function of
$p_{T}$ are consistent with the results from hydrodynamical models,
i.e., the radial flow can push heavier particles to higher $p_{T}$.
Our detailed study helps better understand the chemical and kinetic
properties of the hot and dense QCD matter created in heavy-ion
collisions at LHC energy.

This work was supported in part by the Major State Basic Research
Development Program in China under Contract No. 2014CB845400, the
National Natural Science Foundation of China under contract Nos.
11035009, 11220101005,  11105207, 11275250, U1232206, the Knowledge
Innovation Project of the Chinese Academy of Sciences under Grant
No. KJCX2-EW-N01, and the "Shanghai Pujiang Program" under Grant No.
13PJ1410600.

%\end{CJK*}

\begin{thebibliography}{35}

\bibitem{QCD-QGP} F. Karsch, Nucl. Phys. A \textbf{698}, 199c (2002).

\bibitem{RHICWithePaper} I. Arsene \textit{et al.} (BRAHMS Collaboration), Nucl. Phys. A \textbf{757}, 1
(2005); B. B. Back \textit{et al.} (PHOBOS Collaboration),
\textit{ibid}. A \textbf{757}, 28 (2005); J. Adames \textit{et al.}
(STAR Collaboration), \textit{ibid}. A \textbf{757}, 102 (2005); S.
S. Adler \textit{et al.} (PHENIX Collaboration), \textit{ibid}. A
\textbf{757}, 184 (2005).

\bibitem{RHIC-SQGP}  J. Adams \textit{et al.} (STAR Collaboration), Phys. Rev. Lett. \textbf{95}, 122301 (2005); J. Adams \textit{et al.} (STAR Collaboration), Phys. Rev. Lett. \textbf{93}, 252301 (2004); J. Adams \textit{et al.} (STAR Collaboration), Phys. Lett. B \textbf{612}, 181 (2005); S. S. Adler et al. (PHENIX Collaboration), Phys. Rev. Lett. \textbf{96}, 012304 (2006); L. Adamczyk \textit{et al.} (STAR Collaboration), Phys. Rev. C \textbf{86}, 054908 (2012).

\bibitem{RHIC-pLiquid} P. Huovinen, P. Ruuskanen, Ann. Rev. Nucl. Part. Sci. \textbf{56}, 163 (2006); B. M\"uller, J. L. Nagle, Ann. Rev. Nucl. Part. Sci. \textbf{56}, 93 (2006); U. Heinz and R. Snellings,  Ann. Rev. Nucl. Part. Sci. \textbf{63}, 123 (2013); C. M. Ko {\it et al.}, Nucl. Sci.  Tech. {\bf 24},  050525  (2013);  J. Xu {\it et al.}, Phys. Rev. Lett. {\bf 112}, 012301 (2014); F. M. Liu, Nucl. Sci.  Tech. {\bf 24},  050524  (2013).

\bibitem{ALICE-chDen} K. Aamodt \textit{et al.} (ALICE Collaboration), Phys. Rev. Lett. \textbf{105}, 252301 (2010).
\bibitem{ALICE-chCDepen} K. Aamodt \textit{et al.} (ALICE Collaboration), Phys. Rev. Lett. \textbf{106}, 032301 (2011).
\bibitem{ALICE-chRAA} K. Aamodt \textit{et al.} (ALICE Collaboration), Phys. Lett. B \textbf{696}, 30 (2011).
\bibitem{ALICE-chPCDepen} B. Abelev \textit{et al.} (ALICE Collaboration), Phys. Rev. C \textbf{88}, 044910 (2013).


\bibitem{VISH2} H. Song, S. A. Bass,  U. Heinz, Phys. Rev. C \textbf{83}, 024912 (2011).
\bibitem{HKM} Iu. A. Karpenko, Yu. M. Sinyukov,  K. Werner, Phys. Rev. C \textbf{87}, 024914 (2013); Yu. A. Karpenko,  Yu. M. Sinyukov,  J. Phys. G \textbf{38}, 124059 (2011).

\bibitem{EPOS} K. Werner, I. Karpenko, M. Bleicher, T. Pierog, S. Porteboeuf-Houssais, Phys. Rev. C \textbf{85}, 064907 (2012).

\bibitem{AMPT-LHC}  J. Xu, C. M. Ko, Phys. Rev. C \textbf{83}, 034904 (2011); S. Pal and M. Bleicher, Phys. Lett. B \textbf{709}, 82 (2012).

\bibitem{Thermal-PMB} P. Braun-Munzinger, K. Redlich,  J. Stachel, arXiv:nucl-th/0304013v1; A. Andronic, P. Braun-Munzinger and J. Stachel, Phys. Lett. B \textbf{673}, 142 (2009).

\bibitem{Thermal-EQ} P. Braun-Munzinger, J. Stachel, J. P. Wessels, N. Xu, Phys. Lett. B \textbf{344}, 43 (1995); P. Braun-Munzinger, J. Stachel, J. P. Wessels,  N. Xu, Phys. Lett. B \textbf{365}, 1 (1996); P. Braun-Munzinger, I. Heppe,  J. Stachel, Phys. Lett. B \textbf{465}, 15 (1999); N. Xu, M. Kaneta, Nucl. Phys. A \textbf{698}, 306 (2002).

\bibitem{BLWave} E. Schnedermann, J. Sollfrank,  U. Heinz, Phys. Rev. C \textbf{48}, 2462 (1993).

\bibitem{STARSYS-Spectra} B. I. Abelev \textit{et al.} (STAR Collaboration), Phys. Rev. C \textbf{79}, 034909 (2009).

\bibitem{BLWave-Fabrice} F. Reti\`ere, M. A. Lisa, Phys. Rev. C \textbf{70}, 044907 (2004).

\bibitem{DRAGON} B. Tom\'a\v sik, Comput. Phys. Commun. \textbf{180}, 1642 (2009).
\bibitem{THERMINATOR} M. Chojnacki, A. Kisiel, W. Florkowski, W. Broniowski, Comput. Phys. Commun. \textbf{183}, 746 (2012); A. Kisiel, T. Ta\l u\'c, W. Broniowski,  W. Florkowski, Comput. Phys. Commun. \textbf{174}, 669 (2006).

\bibitem{STAR-HBT-BLWave2005} J. Adams et al. (STAR Collaboration), Phys. Rev. C \textbf{71}, 044906 (2005).

\bibitem{ALICE-Muller} B. M\"uller, J. Schukraft, B. Wys\l ouch, arXiv:1202.3233v1.
\bibitem{ALICE-IP-pp-PbPb} M. Floris (for the ALICE Collaboration), J. Phys. G: Nucl. Part. Phys. \textbf{38}, 124025 (2011).
\bibitem{ALICE-STRANGE-HotQ} D. D. Chinellato (for the ALICE Collaboration), arXiv:1211.7298v1.

\bibitem{LHC-pp} K. Aamodt et al. (ALICE Collaboration), Eur. Phys. J C \textbf{71}, 1 (2011).

\bibitem{Hydro-Flow-RHIC-LHC} P. Bozek, arXiv:1111.4398.

\bibitem{ALICE-raioToPion} R. Preghenella (ALICE Collaboration), arXiv:1203.5904.


\end{thebibliography}
\end{document}